\documentclass[aps,prc,superscriptaddress,reprint]{revtex4-1}
\usepackage{graphicx} 
\usepackage{epstopdf}
\usepackage{dcolumn} 
\usepackage{bm}
\usepackage{hyperref}
\usepackage{array}
\usepackage{setspace}
\usepackage{booktabs}
\hypersetup{
    colorlinks=true,
    linkcolor=blue,
    filecolor=gray,
    urlcolor=blue,
    citecolor=blue,
}
\begin{document}
\title{Rare Isotope Formation in Complete Fusion and Multinucleon Transfer Reactions in Collisions of $^{48}$Ca +$^{248}$Cm around Coulomb Barrier Energies}
\author{Peng-Hui Chen}
\email{Corresponding author: chenpenghui@yzu.edu.cn}
\affiliation{College of Physics Science and Technology, Yangzhou University, Yangzhou 225009, China}
\affiliation{Institute of Modern Physics, Chinese Academy of Sciences, Lanzhou 730000, China}
\author{Fei Niu}
\affiliation{School of Physics and Optoelectronics, South China University of Technology, Guangzhou 510641, China}

\author{Xin-Xing Xu}
\affiliation{Institute of Modern Physics, Chinese Academy of Sciences, Lanzhou 730000, China}


\author{Zu-Xing Yang}
\affiliation{RIKEN Nishina Center, Wako, Saitama 351-0198, Japan}

\author{Xiang-Hua Zeng}
\affiliation{College of Electrical, Power and Energy Engineering, Yangzhou University, Yangzhou 225009, China }

\author{Zhao-Qing Feng}
\email{Corresponding author: fengzhq@scut.edu.cn}
\affiliation{School of Physics and Optoelectronics, South China University of Technology, Guangzhou 510641, China}

\date{\today}
\begin{abstract}
  Within the framework of dinuclear system model, the reaction mechanisms for synthesizing target-like isotopes from Bk to compound nuclei Lv are thoroughly investigated in complete and incomplete fusion reaction of  $^{48}$Ca +$^{248}$Cm around Coulomb barrier energies. Production cross section of $^{292,293}$Lv as a function of excitation energy in fusion-evaporation reactions and target-like isotopic yields in multinucleon transfer reactions are evaluated, in which a statistical approach is used to describe the decay process of excited nuclei.
The available experimental data can be reproduced well with the model reasonably.
The products of all possible formed isotopes in the dynamical pre-equilibrium process for collision partners at incident energy $E_{\rm lab}$ = 5.5 MeV/nucleon are exported, systematically.
It is found that the quasi-fission fragments are dominant in the yields. The optimal pathway from target to compound nuclei shows up along the valley of potential surface energy. 
The effective impact parameter of two colliding partners leading to compound nuclei is selected from head on collison to semicentral collision with $L$ = 52 $\rm \hbar$. The timescale boundary between complete fusion and multinucleon transfer reactions is about 5.7$\times 10^{-21}$ s with effective impact parameters. Synthesis cross section of unknown neutron-rich actinides from Bk to Rf have been predicted around several nanobarn.

\begin{description}
\item[PACS number(s)]
25.70.Jj, 24.10.-i, 25.60.Pj
\end{description}
\end{abstract}

\maketitle

\section{Introduction}

To find the limits of the nuclear landscape, theoretical and experimental nuclear physicists devote to explore synthesis of exotic nuclei and superheavy elements (SHE) toward drip lines and island of stability via heavy ion collisions. For producing unknown superheavy elements, fusion-evaporation reactions have been widely used in different laboratories all over the world. In reactions of actinides with a double magic $^{48}$Ca beam at Flerov Laboratory of Nuclear Reactions (FLNR, Dubna) \cite{og15,ut15}, the synthesis of SHEs with atomic number Z up to 118 has been claimed. In Gesellschaft f$\ddot{ \rm u}$r Schwerionenforschung (GSI), the production of superheavy elements with Z = 107-112,114-117 has been identified\cite{ho11,mu15}. The production of new element Nihonium (Z = 113) in collision of $^{70}$Zn + $^{209}$Bi has been observed at RIKEN\cite{mo15}. The element Flerovium (Z=114) has been sythesized in Lawrence Berkeley National Laboratory (LBNL, Berkeley)\cite{st09}. Chinese group SHE has sythesized superheavy isotopes Dubnium (Z = 105), Bohrium (Z=107), Darmstadtium (Z=110) in Institute of Modern Physics(IMP, Lanzhou)\cite{ga01,ga04,zh12}. To produce exotic transuranium isotopes, multinucleon transfer (MNT) reactions are proposed to perform experiments with radioactive beams in laboratories. It has advantages that products are formed with wide mass region owing to broad excitation function in the MNT products. The complete fusion reactions between two heavy partners at energies around the Coulomb barrier are strongly damped by competing incomplete fusion reactions (quasi-fission and deep-inelastic reactions). Therefore, more insightful theoretical and experimental studies of the reaction mechanisms are required to make precise prediction for the probability of compound nuclei and MNT products in such reactions.

The quasi-fission and deep-inelastic heavy-ion collisions have been extensively investigated in experiments since 1970s, in which MNT reactions had been proposed to synthesize superheavy elements initially. However, new neutron-rich projectile-like fragments and proton-rich actinide nuclei were observed in laboratories \cite{Art71, Art73, Art74, Hil77, Gla79, Moo86, Wel87}. In particular, isospin asymmetric collisions may provide valuable information on production mechanism of exotic heavy nuclei. In laboratories worldwide, the reactions of $^{136}$Xe+$^{208}$Pb \cite{Ko12,Ba15}, $^{136}$Xe+$^{198}$Pt \cite{Wa15}, $^{156,160}$Gd+$^{186}$W \cite{Ko17}, $^{238}$U+$^{232}$Th \cite{Wu18} have been performed to creat unknown neutron-rich heavy nuclei nearby neutron shell N = 126, applied to understand the origin of heavy elements in nuclear astrophysics. 

Following the motivation for predicting exotic heavy and superheavy nuclei, several models have been developed, such as the dynamical model based on multidimensional Langevin equations \cite{Za07,Za08}, the time-dependent Hartree-Fock (TDHF) approach \cite{Gola09,Seki16,Gu18,Ji18}, the GRAZING model \cite{Wint94,bib:6}, the improved quantum molecular dynamics (ImQMD) model \cite{Zha15,li16}, the Langevin-type dynamical equations \cite{sh11,bo11}, and the dinuclear system (DNS) model\cite{Fe09,bib:8,ch20,bao18,ba21,zhu21}, etc. Some interesting issues of synthesis mechanism, total kinetic energy spectra, structure effect have been stressed. There are still some open problems for strongly damped reactions, for example, the mechanism of pre-equilibrium particles emission, the stiffness of nuclear surface during the nucleon transfer process, the mass limitation of new isotopes with stable heavy target nuclides, etc.

In the three laboratories, FLNR\cite{og00}, GSI\cite{ho12} and RIKEN\cite{ka17} obtain cross-section excitation functions of 3n, 4n evaporation channels for production of superheavy element Z=116 in $^{48}$Ca induced reactions with $^{248}$Cm targets, early or late, respectively. In the experiments of synthesizing superheavy nuclei (Z=116) with $^{48}$Ca + $^{248}$Cm \cite{Deva15,hof85,gag86,hez16,dev19}, the massive independent yields of target-like fragments have been observed, especially new heavy isotopes involved. Production cross section of all formed products brings us an opportunity to investigate the interplay between equilibrium and dissipation for low energy heavy-ion collisions as well as decay properties of excited SHN.
So it attracts our interests to explore nuclear dynamics of reaction mechansims in complete and incomplete fusion in terms of evolution time and dissipation energy.

In this work, the $^{48}$Ca induced complete and incomplete fusion reactions with the combination of $^{248}$Cm are calculated with the DNS model.
The aim of this paper is to study the dynamics on the sythesis cross sections of nuclides in complete and incomplete fusion of $^{248}$Cm with $^{48}$Ca projectile.
The article is organized as follows: In Sec. \ref{sec2} we give a brief description of the DNS model. Calculated results and discussions are presented in Sec. \ref{sec3}. Summary is concluded in Sec. \ref{sec4}.

\section{Model description}\label{sec2}

The dynamical complete and incomplete fusion mechanisms are described as a diffusion process, in which resulting distribution probability is obtained by solving a set of master equations numerically in the potential energy surface of the DNS. The time evolution of the distribution probability $P(Z_{1},N_{1},E_{1},\beta, t)$ for fragment 1 with proton number $Z_{1}$, neutron number $N_{1}$, excitation energy $E_{1}$, quadrupole deformation $\beta$ is described by the following master equations:
\begin{eqnarray}
\label{mst}
&&\frac{d P(Z_1,N_1,E_1,\beta,t)}{d t} =  \nonumber \\ 
&&  \sum \limits_{Z^{'}_1}  W_{Z_1,N_1,\beta;Z'_1,N_1,\beta}(t) [d_{Z_1,N_1} P(Z'_1,N_1,E'_1,\beta,t) \nonumber   \\
&& - d_{Z'_1,N_1}P(Z_1,N_1,E_1,\beta,t)]   \nonumber   \\
&& + \sum \limits_{N'_1}  W_{Z_1,N_1,\beta;Z_1,N'_1,\beta}(t) [d_{Z_1,N_1}P(Z_1,N'_1,E'_1,\beta,t) \nonumber   \\ 
&& - d_{Z_1,N'_1}P(Z_1,N_1,E_1,\beta,t)]
\end{eqnarray} 
The $W_{Z_{1},N_{1},\beta;Z^{'}_{1},N_{1},\beta}$($W_{Z_{1},N_{1},\beta,;Z_{1},N^{'}_{1},\beta}$) is the mean transition probability from the channel($Z_{1},N_{1},E_{1},\beta$) to ($Z^{'}_{1},N_{1},E^{'}_{1},\beta$), [or ($Z_{1},N_{1},E_{1},\beta$) to ($Z_{1},N^{'}_{1},E^{'}_{1},\beta$)], and $d_{Z_{1},Z_{1}}$ denotes the microscopic dimension corresponding to the macroscopic state ($Z_{1},N_{1},E_{1}$).The sum is taken over all possible proton and neutron numbers that fragment ($Z^{'}_{1}$,$N^{'}_{1}$) may take, but only one nucleon transfer is considered in the model with the relations $Z^{'}_{1}=Z_{1}\pm1$ and $N^{'}_{1}=N_{1}\pm1$. 

The motion of nucleons in the interacting potential is governed by the single-particle Hamiltonian. The excited DNS opens a valence space in which the valence nucleons have a symmetrical distribution around the Fermi surface. Only the particles at the states within the valence space are actively for nucleon transfer. The transition probability is related to the local excitation energy and nucleon transfer, which is microscopically derived from the interaction potential in valence space as described in \cite{Ch17,No75}.


The local excitation energy is determined by the dissipation energy from the relative motion and the potential energy surface of the DNS as
\begin{eqnarray}
\label{lee}
\varepsilon^{\ast}(t)=E^{\rm diss}(t)-\left(U(\{\alpha\})-U(\{\alpha_{\rm EN}\})\right).
\end{eqnarray}
The entrance channel quantities $\{\alpha_{\rm EN}\}$ include the proton and neutron numbers, quadrupole deformation parameters and orientation angles being $Z_{ \rm P}$, $N_{\rm P}$, $Z_{\rm T}$, $N_{\rm T}$, $R$, $\beta_{\rm P}$, $\beta_{\rm T}$, $\theta_{\rm P}$, $\theta_{\rm T}$ for projectile and target nuclei with the symbols of $P$ and $T$, respectively. The interaction time $\tau_{\rm int}$ is obtained from the deflection function method \cite{Wo78}. The energy dissipated into the DNS increase exponentially \cite{Fe07b}.
The potential energy surface (PES) of the DNS is evaluated by
\begin{eqnarray}\label{dri}
U_{\rm dr}(t) = Q_{\rm gg}+V_{\rm C}(Z_1,N_1,\beta_1,Z_2,N_2,\beta_2,t)     \nonumber   \\ 
+ V_{\rm N}(Z_1,N_1,\beta_1,Z_2,N_2,\beta_2,t) + V_{\rm def}(t)
\end{eqnarray}
which satisfies the relation of $ Z_{1}+Z_{2}=Z $ and  $ N_{1}+N_{2}=N$ with the $Z$ and $N$ being the proton and neutron numbers of composite system, respectively. 
The $B(Z_{\rm i},N_{\rm i}) (i=1,2)$ and $B(Z,N)$ are the negative binding energies of the fragment $(Z_{\rm i},N_{\rm i})$ and the composite system $(Z,N)$, respectively. The $\theta_{i}$ denotes the angles between the collision orientations and the symmetry axes of the deformed nuclei. 
$V_{\rm def}(t)$ is the deformation energy of DNS at the moment $t$.
The evolutions of quadrupole deformations of projectile-like and target-like fragments undergo from the initial configuration as
\begin{eqnarray}\label{qde}
\beta' _{\rm T} (t) = \beta _ {\rm T} \exp(-t/\tau_{\rm \beta}) + \beta_1 [ 1 - \exp(-t/\tau_{\rm \beta})],      \nonumber \\
\beta' _{\rm P} (t) = \beta _ {\rm P} \exp{(-t/\tau_{\rm \beta})} + \beta_2 [ 1 - \exp(-t/\tau_{\rm \beta})]
\end{eqnarray}
with the deformation relaxation is $\tau_{\rm \beta}=4\times10^{-21} \ s$.

The total kinetic energy (TKE) of the primary fragment is evaluated by
\begin{equation}\label{tke}
 TKE (A_{1}) = E_{\rm c.m.} + Q_{ \rm gg}(A_{1}) - E^{\rm diss}(A_{1}),
\end{equation}
where $Q_{\rm gg} = M_{\rm P} + M_{\rm T} - M_{\rm PLF} -M_{\rm TLF}$ and $E_{\rm c.m.}$ being the incident energy in the center of mass frame. The mass $M_{\rm P}$, $M_{\rm T}$, $M_{\rm PLF}$ and $M_{\rm TLF}$ correspond to projectile, target, projectile-like fragment and target-like fragment, respectively.


The cross-sections of the survival fragments produced in MNT reactions and fusion-evaporation residue cross-sections are evaluated by
\begin{eqnarray}\label{mnt}
&&\sigma_{\rm sur}(Z_{1},N_{1},E_{\rm c.m.})= \frac{\pi \hbar^{2}}{2\mu  E_{\rm c.m.}}\sum_{J=0}^{J_{\max}} (2J+1)     \nonumber \\
&& \times  \int  f(B) T(E_{\rm c.m.},J,B) \sum_{s}P(Z^{'}_{1},N^{'}_{1},E^{'}_{1},J^{'}_{1},B)   \nonumber \\
&& \times  W_{\rm sur}(Z^{'}_{1},N^{'}_{1},E^{'}_{1},J^{'}_{1},s)dB
\end{eqnarray}
and
\begin{eqnarray}\label{ffr}
\sigma^s_{\rm ER}(E_{\rm c.m.}) =&&\frac{\pi\hbar^2}{2\mu E_{c.m.}} \sum \limits ^{J_{max}}_{J=0} (2J+1)T(E_{\rm c.m.},J) \nonumber  \\  && \times P_{\rm CN}(E_{\rm c.m.},J) W^{\rm s}_{ \rm sur}(E_{\rm c.m.},J),
\end{eqnarray}
respectively. The $\mu$ is the reduced mass of relative motion. 
The transmission probability $T(E_{\rm c.m.},J)$ is calculated by Hill-Wheeler formular in combination with barrier distribution function.
The $E_{1}$ and $J_{1}$ are the excitation energy and the angular momentum for the fragment (Z$_{1}$, N$_{1}$). The maximal angular momentum $J_{\max}$ is taken to be the grazing collision of two nuclei. The survival probability $W_{\rm sur}$ of each fragment is evaluated with a statistical approach based on the Weisskopf evaporation theory \cite{Ch16}, in which the excited primary fragments are cooled in evaporation channels $s(Z_{s},N_{s})$ by $\gamma$-rays, light particles (neutrons, protons, $\alpha$ etc) in competition with the binary fission via $Z_{1}=Z^{'}_{1}-Z_{s}$ and $N_{1}=N^{'}_{1}-N_{s}$. The $P_{\rm CN}(E_{\rm c.m.},J)$ are fusion probability which sum over all the fragments probability located outside of BG (Businaro Gallone) point. The transfer cross-section is smoothed with the barrier distribution. 


\section{Results and discussion}\label{sec3}

\begin{figure*}[htb]
\includegraphics[width=.85\linewidth]{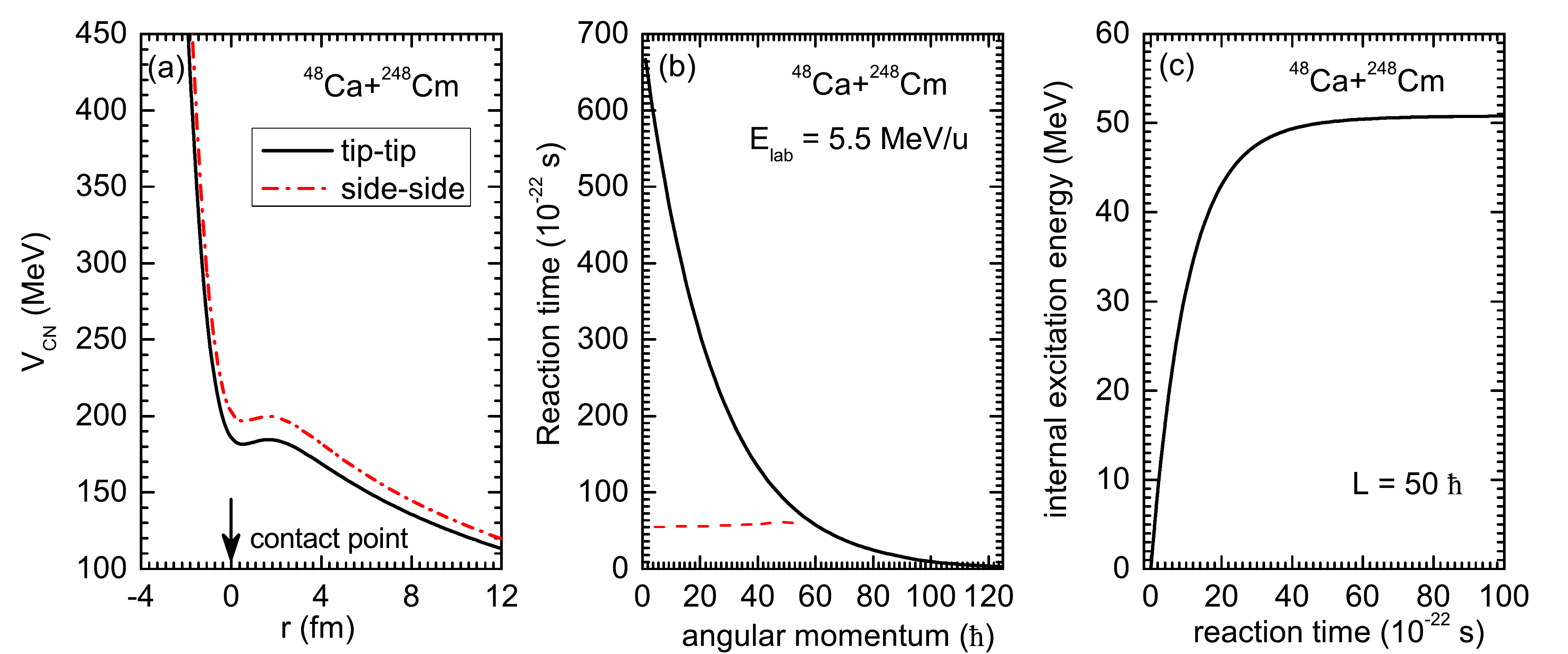}
\caption{\label{fig1} (Color online) Solid black line and red dash-dots line are interaction potential of the tip-tip and side-side collisions as a function of surface distance in the reaction of $^{48}$Ca+$^{248}$Cm in panel (a); In panel (b), Solid black and red dash lines corresponding to the reaction the "boundary" of timescale between complete fusion and MNT reactions; The internal excitation energy of $^{48}$Ca+$^{248}$Cm collision at $E_{\rm lab}$ = 5.5 MeV/nucleon with impact parameter $L$ = 50 $\hbar$ in panel (c).}
\end{figure*}
In heavy-ion collisions, overcoming Coulomb barrier, kinetic energy of relative motion transforms rapidly into internal excitation of dinuclear system at contact point.
The interaction potential distribution to distance, interaction time to impact parameter and internal excitation energy to reaction time for the systems of $^{48}$Ca + $^{248}$Cm reactions at incident energy $E_{\rm lab}$ = 5.5 MeV/nucleon are shown in Fig.\ref{fig1}. The interaction potential is calculated as a function of surface distance bewteen two heavy partners.
From panel (a), one can see that Coulomb barrier of colliding system is about 185 MeV, quasi-fission barrier is several MeV. The potential pocket locates nearby contact point almost. 
The reaction time is calculated by deflection function, plotting as a function of angular momentim, which decreases exponentially with increasing angular momentum. 
The internal excitation energy dissipating in dinuclear system increases exponentially with increasing evolution time. 
The existence of the pocket in the entrance channel is crucial for the compound nucleus formation in fusion reactions, which is the input physical quantity in calculating capture cross section. 
The barrier is taken as the potential value at the touching configuration and the nucleus-nucleus potential is calculated with the same approach in fusion reactions \cite{Fe06}. 
According to Fig. \ref{fig1}, it was found that there are few MeV potential pocket for the heavy systems, because of the strong Coulomb repulsion between two colliding partners with $Z_1 Z_2 = 1860 $. The lighter collision systems own the deeper potential pocket relatively. The deeper potential pocket collision system leads to the longer reaction time correspondingly. With impact parameter $L$ = 50 $\hbar$ for $^{48}$Ca+$^{248}$Cm reaction, the timescale of reaching the almost equilibrium state for incident energy dissipating in internal excitation energy is about $5.7 \times 10^{-21}$s.

\begin{figure}[htb]
\includegraphics[width=1.\linewidth]{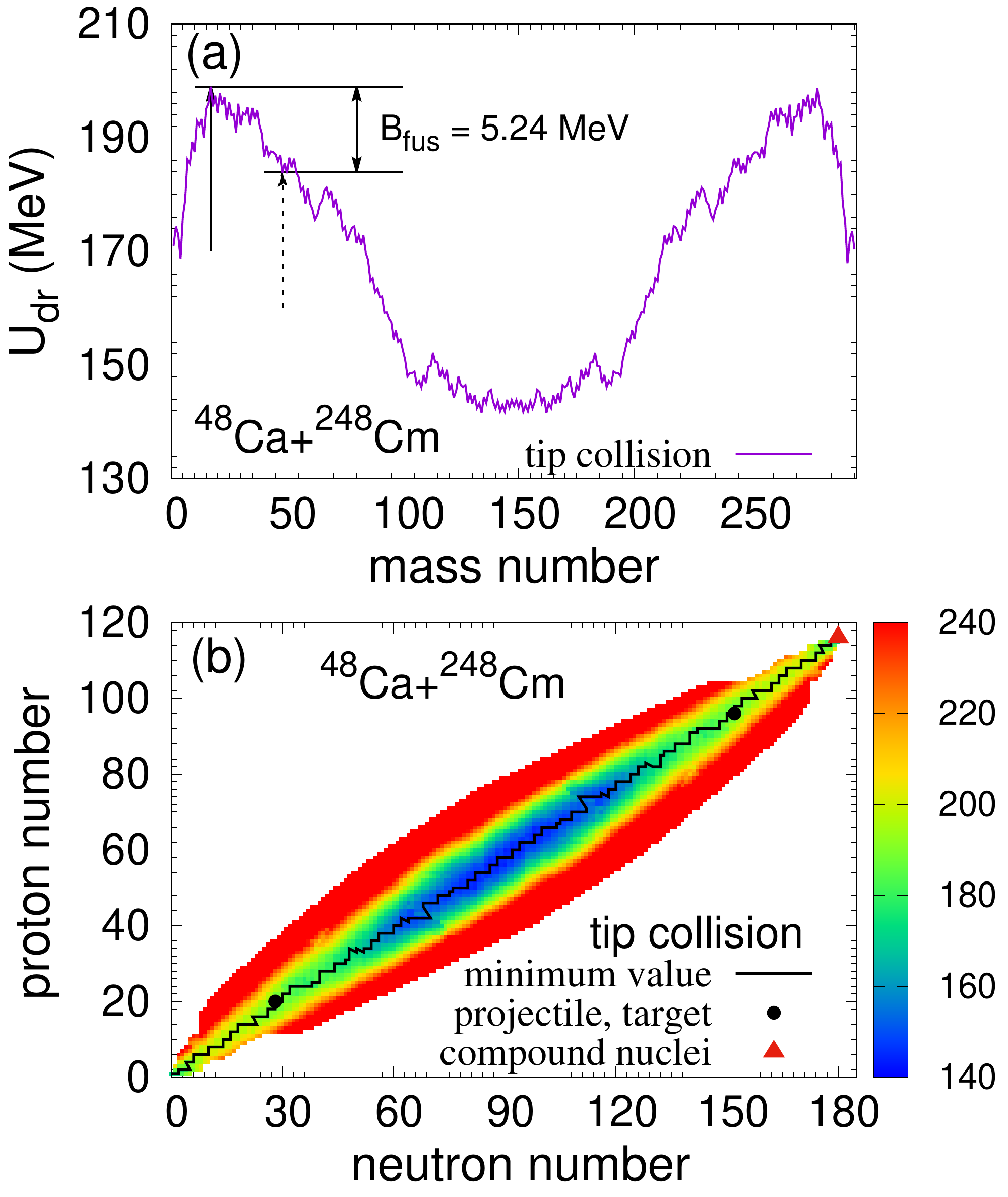}
\caption{\label{fig2} (Color online) The driver potentials of $^{48}$Ca+$^{248}$Cm at tip-tip collision. Potential energy surface as a function of mass calculated at two fixed distances between projectile and target. The arrows and black solid circles indicates the entrance channel. Solid triangle is the compound nuclei. The two solid lines are minimum value in two-dimensions potential energy surface.}
\end{figure}

The nucleons can be transferred between the collision partners resulting in the internal degree of freedom charactering the nuclear states encounter a rapid rearrangement along the potential energy surface (PES) as well as dissipating their kinetic energy and angular momentum. The calculation of multi-dimensional adiabatic PES for heavy nuclear system is a quite complicated physical problem, which is still an open problem, so far. In this work, PES for tip-tip collisions of $^{48}$Ca+$^{248}$Cm is calculated by Eq. (\ref{dri}) as a diabatic type with frozen distance, shown in Fig. \ref{fig2} (b).
The solid black line, solid black circle and solid red triangle are valley value, projectile-target position and compound nuclei, respectively. 
The valley value in PES is listed as a function of mass, shown in Fig. \ref{fig2} (a).
The inner fusion barrier of the collision partners is $B_{\rm fus}$ = 5.24 MeV, which means that it needs to overcome 5.24 MeV barrier energy to fuse.
The DNS fragments towards the mass symmetric valley release the positive energy, which is available for nucleon transfer. The spectra exhibits a symmetric distribution for each isotopic chain. The valley in the PES is close to the $\beta$-stability line and enable the diffusion of the fragment probability.

\begin{figure}[htb]
\includegraphics[width=1.\linewidth]{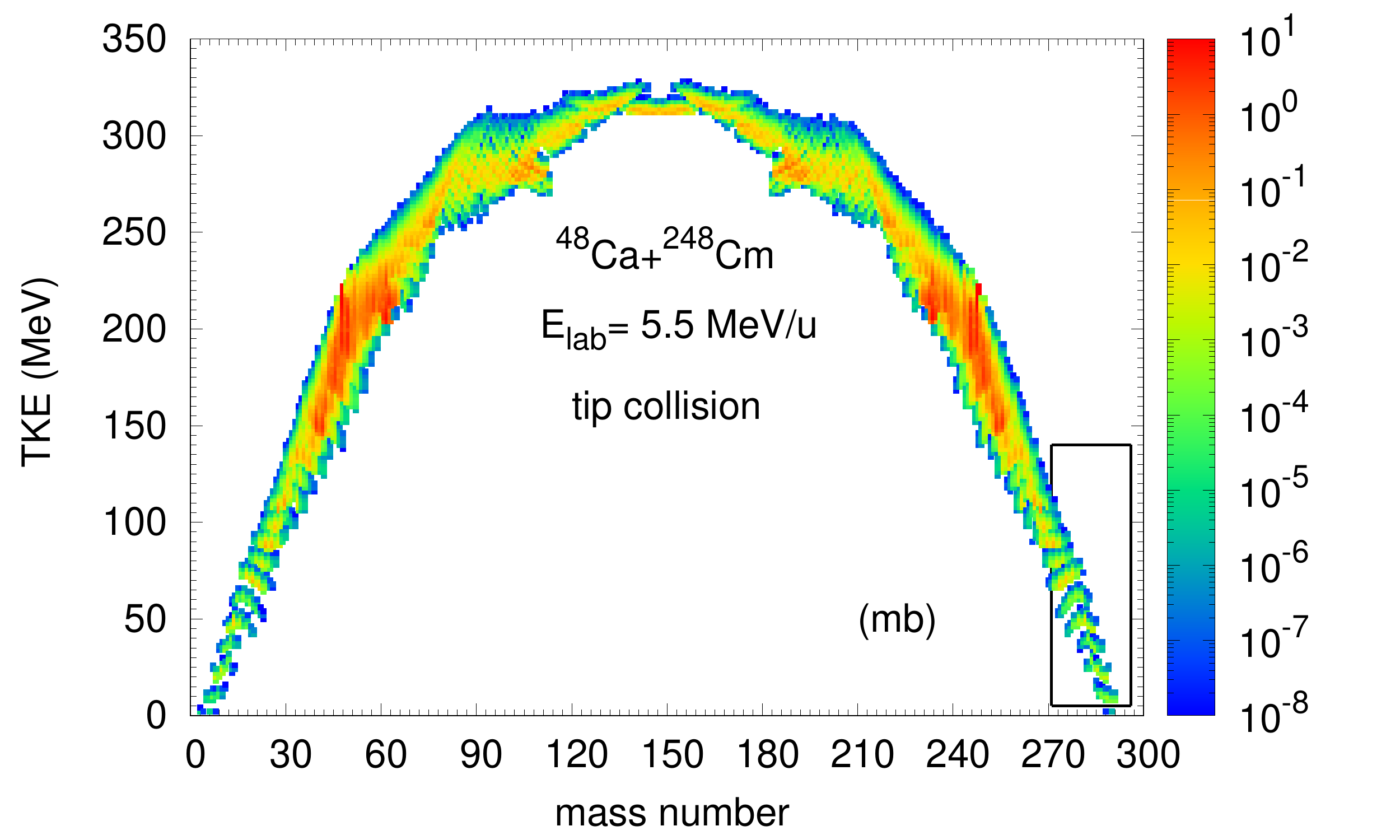}
\caption{\label{fig3}(Color online) Calculated TKE-mass distribution of primary reaction products in the collision of $^{48}$Ca+$^{248}$Cm at $E_{\rm lab}$ = 5.5 MeV/nucleon. The fragments in square areas are overcome the inner fusion barrier (Businaro Gallone point). }
\end{figure}
Figture \ref{fig3} shows the calculation correlation of the total kinetic energy (TKE) and mass distributions of the reaction products along with inclusion mass distribution for the $^{48}$Ca+$^{248}$Cm reaction at near-barrier energy of $E_{\rm lab}$ = 5.5 MeV/nucleon. This calculations agree roughly well with experimental data \cite{itk04}.
They are consistent with that calculated by Langevin-type dynamical equations \cite{zag05}.
In most of the damped collisions the interaction time is rather short (several units of 10$^{-21}$ s). These fast events correspond to grazing collisions with intermediate impact parameters, which are shown by the areas around projectile-target points. A large amount of kinetic energy is dissipated here very fast at relatively low mass transfer (more than 45 MeV during several units of several units of 10$^{-21}$ s). The other events correspond to much slower collision with large overlap of nuclear surface and significant nucleons rearrangement. In the TKE-mass plot, these events spread over a wide region of mass fragments. These fragments in the square areas indicate overcome inner barrier (Businaro Gallone point), which means they can lead to compound nuclei, in the framework of DNS model. 


\begin{figure}[htb]
\includegraphics[width=1.\linewidth]{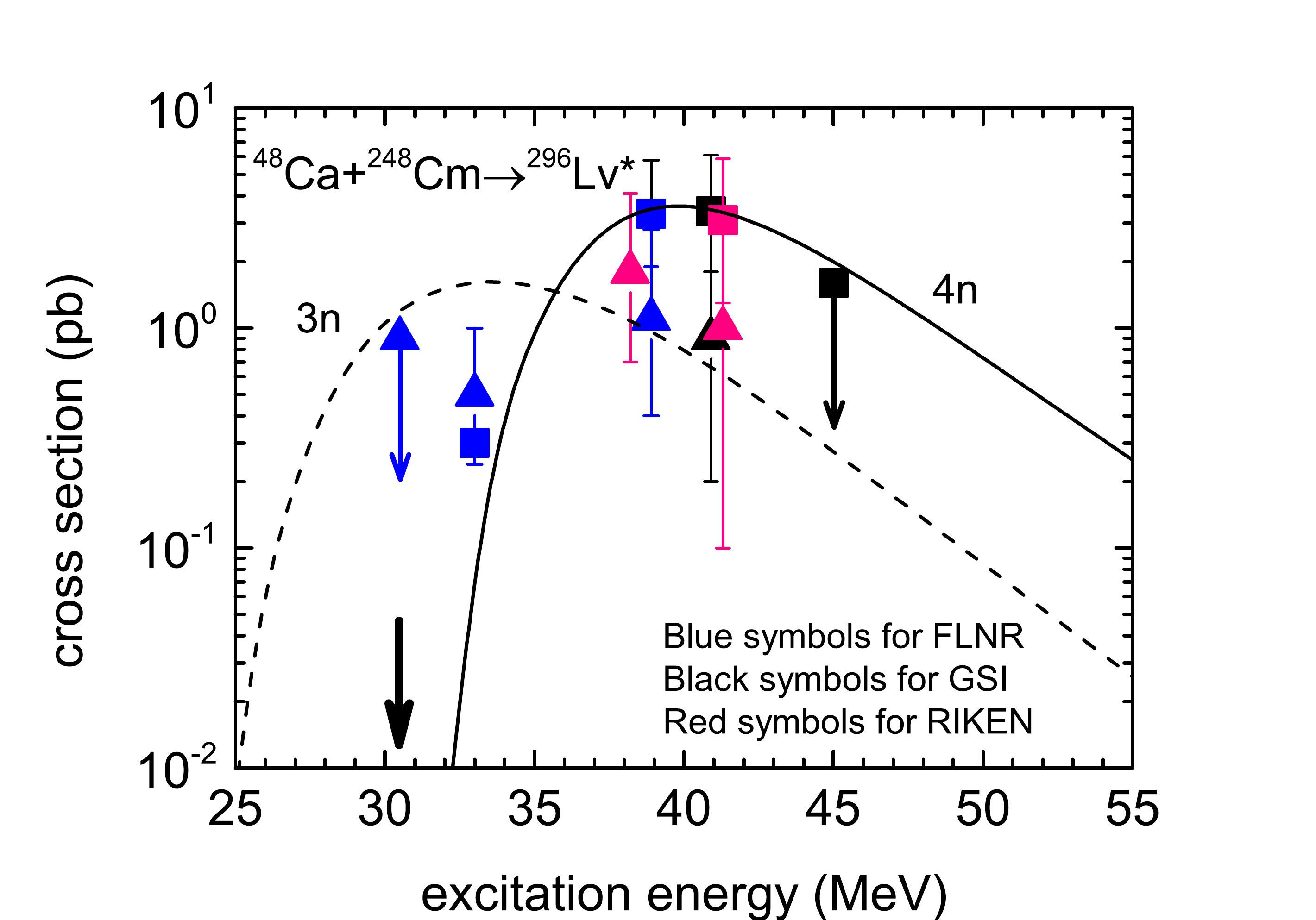}
\caption{\label{fig4}(Color online) The calculated evaporation residues as a function of excitation energy in the reaction $^{48}$Ca($^{248}$Cm, xn). Cross-sections and cross-section limits of the reaction $^{48}$Ca + $^{248}$Cm $\rightarrow$  $^{296}116^*$ measured in GSI\cite{ho12}, FLNR\cite{og00} and RIKEN\cite{ka17}. The data for synthesis of $^{293}$116 (3n channel, triangles) and $^{292}$116 (4n channel, squares) are shown.}
\end{figure}
Predicted and experimental excitation functions of 3n and 4n channels for production of Livermorium (Z=116) in the $^{48}$Ca induced reactions are shown in Fig. \ref{fig4}. 
The experimental data have been obtained at the FLNR, GSI and RIKEN, which correspond to blue symbol, black symbol and red symbol. 
Early or late, FLNR\cite{og00}, GSI\cite{ho12} and RIKEN\cite{ka17} obtain excitation functions of 3n, 4n evaporation channels for production of superheavy element with Z=116 in collision of $^{48}$Ca + $^{248}$Cm, respectively.
In FLNR laboratory, at lower beam energies three irradiations were performed in Dubna in June-July and November-December 2000 and January and April-May 2001 \cite{og00}.
At $E^*$ = 30.5 MeV a cross-section limit of 0.9 pb was reached. At $E^*$ = 33.0 MeV, three decay chains were measured resulting in a cross-section of ($0.5^{+0.5}_{-0.26}$) pb, which were
assigned to $^{293}$ 116. At the same excitation energy a cross-section limit of 0.3 pb was obtained for the 4n channel.
The highest energy studied resulted in $E^*$ = 38.9 MeV. At this energy, six decay chains were measured and assigned
to $^{292}$116 resulting in a cross-section of ($3.3^{+2.5}_{-1.4}$) pb for the 4n channel. 
Also at the same energy two chains from the 3n channel were measured resulting in a cross-section of ($1.1^{+1.7}_{-0.7}$) pb. 
This experiment was performed in April-May, 2004 \cite{oga04}.
Four cross sections data marked by solid blue symbols listed in Fig. \ref{fig4}.

In GSI laboratary, at energy $E^{* }$ = 40.9 MeV, they detected six decay chains, four events were assigned to the 4n channel resulting in a cross section of ($3.4^{+2.7}_{-1.6}$) pb and one from the other two events to the 3n channel. 
Therefore the cross-section of ($0.9^{+2.1}_{-0.7}$) pb is present here, which is valid for the event definitely as signed to $^{293}$116. 
No event was observed in the second part of the experiment at $E^*$ = 45.0 MeV, resulting in a one-event cross-section limit of 1.6 pb.
Three cross sections data marked by solid black symbols list in Fig. \ref{fig4}.

In RIKEN laboratory, the fusion reaction $^{48}$Ca + $^{248}$Cm $\rightarrow$ $^{296}$Lv$^*$ was investigated using the gas-filled recoil ion separator GARIS at
RIKEN. The reaction was studied at excitation energies of 41.3 and 38.2 MeV. A total of seven decay chains were
observed. Three of the chains were assigned to the decay of $^{292}$Lv and three to the decay of $^{293}$Lv.
The resulting cross sections are $\sigma_{4n} = (3.1 ^{+2.8}_{-1.8})$ pb at $E^*$ = 41.3 MeV, $\sigma_{3n} = (1.0 ^{+2.4}_{-0.9})$ pb, and
$\sigma_{3n} = (1.8 ^{+2.3}_{-1.1})$ pb at $E^*$ = 41.3 and 38.2 MeV, respectively. In the case of unobserved decay chains, the one event cross section limits are 1.9 and 1.6 pb at $E^*$ = 41.3 and 38.2 MeV, respectively. 
Three cross-sections data marked by solid red symbols in Fig. \ref{fig4}.
In theory calculation, $Q_{ \rm value}$ of the reactions $^{48}$Ca+$^{248}$Cm $\rightarrow$ $^{296}$Lv$^*$ is -166.57 MeV, the $V_{ \rm Bass}$ potential is 197.12 MeV, which indicated by solid black arrow. 
The dash line and solid line are calculated excitation functions correspond to 3n and 4n evaporation channels. One can see from Fig.\ref{fig4} that calculated excitation functions have a good agreement with  all the available experimental data\cite{ho12,og00,ka17}.

\begin{figure*}[htb]
\includegraphics[width=.9\linewidth]{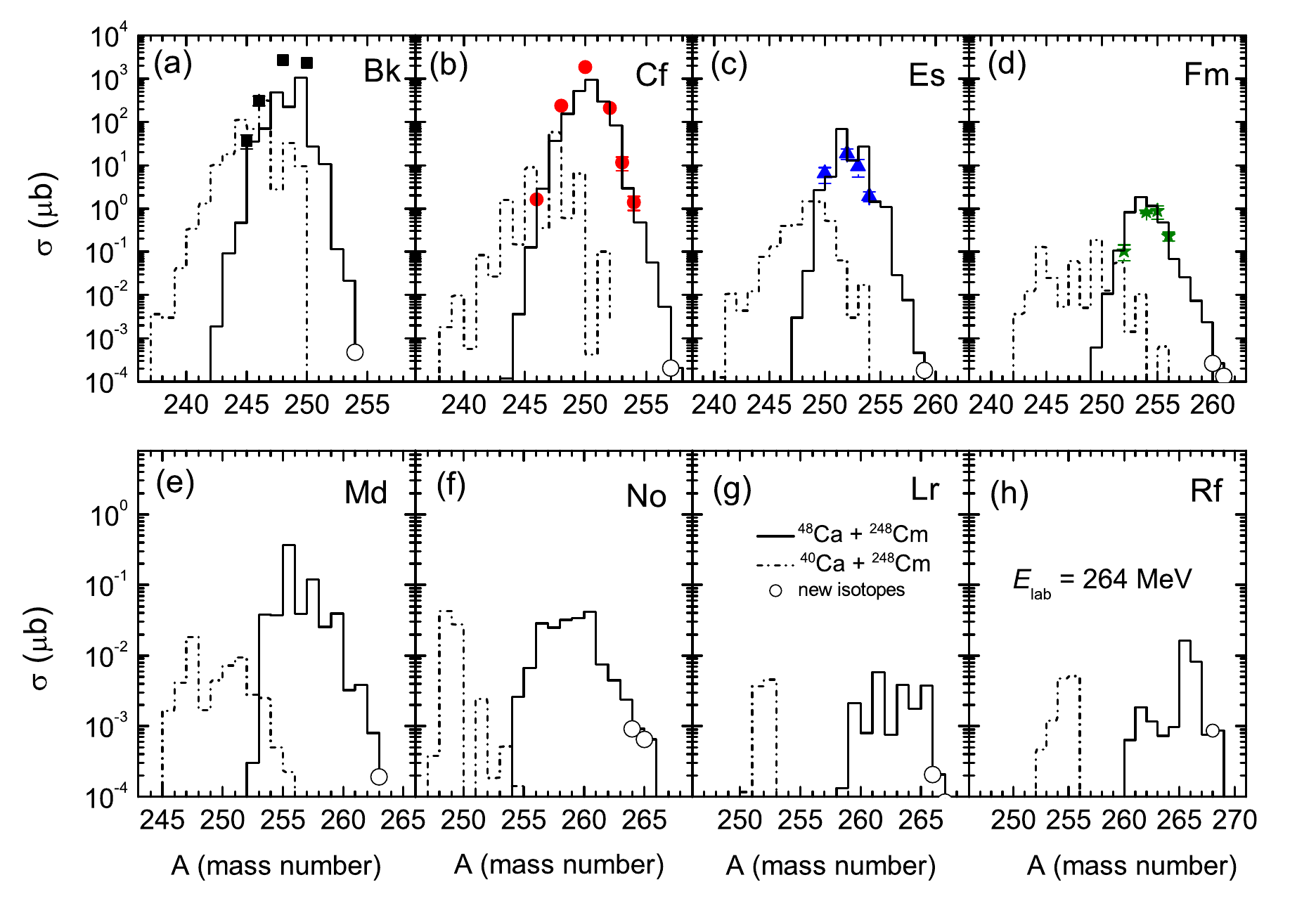}
\caption{\label{fig5}(Color online) Predicted and measured isotopic distribution for the production of target-like fragments with Z = 97-100 in the collisions of $^{48}$Ca+$^{248}$Cm. The experimental data are taken from Gesellschaft f$\ddot{ \rm u}$r Schwerionenforschung (GSI, Germany)\cite{hof85} and Lawrence Berkeley Laboratory (LBL, America)\cite{gag86}.}
\end{figure*}

In collision of $^{48}$Ca+$^{248}$Cm at energies around Coulomb barrier, the MNT products are dominant in all isotopic yields.
In 1980s, to study the role of neutron-rich projectile $^{48}$Ca in enhancing the yields of neutron-rich actinides and to determine what effect the eight fewer neutrons in $^{48}$Ca would have on the mass distribution, two series of experiments were performed at LBL and at GSI using the radiochemical methods and on-line gas-jet transport of short-lived reaction products combined with electronic detection systems\cite{hof85}.
Above-target isotopes from Bk to Fm,  below target isotopes of Rn, Ra, Ac, Th, U, Pu have been observed in the reactions of $^{48}$Ca + $^{248}$Cm at incident energies $E_{\rm lab}$ = 223-239, 248-263, 247-263, 272-288, 304-318 MeV. The max yields of above target isotopes are around $E_{\rm lab}$ = 248-263 MeV. In this paper, we report on production of Bk, Cf, Es and Fm only. The production of below target isotopes have been discussed in paper \cite{zhi21}. 

In the year 2000, experiments of $^{48}$Ca+$^{248}$Cm at incident energies $E_{\rm lab}$ = 265.4, 270.2 MeV have been performed in GSI\cite{dev19}. In the experiment, fusion products and target-like transfer reaction products have been measured on SHIP. Due to short dection time (two days) and limited seperation method, several above target isotopes have been obtained. They are $^{252,254}$ Cf, $^{254,256}$Es, $^{254,256}$Fm listed in TABLE \ref{tab1}. 

The cross-sections measured in ref. \cite{hez16} for the same isotopes and same collision system are also listed for comparison. The results from both experiments are quite well in agreement despite the different experimental techniques and systematic uncertainties. For example, cross-sections for the directly populated nuclides $^{254}$Cf and $^{254}$Es are in agreement within a factor of $\approx$ 1.8 and 1.5 times compared to the cross-sections presented in paper \cite{hez16}. For the same reaction system, different seperation methods can result in several orders magnitude discrepancy. 
Predicted and identified MNT products from Bk to Fm along with the measured cross-sections is presented in Fig. \ref{fig5}. Compared to the measured cross- section, calculation have a good agreement reasonably.
Calculation and experiemntal data reveal a trend that cross section of a certain MNT products decrease on average by one order of magnitude with the transfer of each proton from the projectile to the target nucleus, because of the heavier above-target isotopes and the smaller fission barriers. Here predictions have been made for unknown isotopes $^{254}$Bk, $^{257}$Cf, $^{259}$Es, $^{260}$Fm, $^{263}$Md, $^{264,265}$No, $^{266}$Lr and $^{268}$Rf, which are 0.4, 0.2, 0.1, 0.2, 0.2, 0.9, 0.6, 0.2 and 1 nb respectively. 

\begin{table}[h!]
\caption{Calculated and measured cross sections of isotopes are collected for the reactions of $^{48}$Ca + $^{248}$Cm around the Coulomb barrier energies. 
The errors listed in parentheses behind cross section value.}
\label{tab1}
\begin{spacing}{1.19}
\begin{tabular}{c|cccc}
\hline
Isotope & Exp.($\mu$b)  & Exp($\mu$b) & Exp.($\mu$b) & Cal.($\mu$b)  \\
 (MeV)    & (270)         & (247-263) & (272-288) & (264)  \\
Ref. &   \cite{dev19}   &  \cite{hof85}  &  \cite{hof85}  & this work    \\
\hline
$^{245}$Bk  &                        & 40 $  (40\%)$ & 67  $(20\%)$   & 35.0    \\
$^{246}$Bk  &                        & 360 $ (10\%)$ & 480 $ (15\%)$  & 69.6    \\
$^{248}$Bk  &                        & 2900 $ (6\%)$ & 2680 $ (15\%)$ & 224.0    \\
$^{250}$Bk  &                        & 2520 $(5\%)$   & 2920 $ (10\%)$ & 1061.0    \\
\hline
$^{246}$Cf  &                        & 1.8  $ (2\%)$ & 1 $ (20\%)$    & 2.8    \\
$^{248}$Cf  &                        & 260  $ (2\%)$   & 210 $ (6\%)$   & 151.0    \\
$^{250}$Cf  &                        & 2380 $ (7\%)$  & 1935 $ (2\%)$  & 930.0    \\
$^{252}$Cf  &$\textless 58.0$ $\mu$b          & 225  $ (4\%)$   & 220 $ (15\%)$  & 81.0    \\
$^{253}$Cf  &                        & 12  $ (20\%)$   & 4 $ (15\%)$    & 2.9    \\
$^{254}$Cf  &$\textgreater 0.12$ $\mu$b     & 1.5 $ (30\%)$  & 1 $ (25\%)$    & 0.5    \\
 \hline
$^{250}$Es  &                        & 6.6 $ (40\%)$  &                     & 5.5    \\
$^{252}$Es  &                        & 30 $ (15\%)$   & 24 $ (5\%)$    & 12.0    \\
$^{253}$Es  &                        & 10 $ (10\%)$   & 7.8 $ (15\%)$  & 26.0    \\
$^{254}$Es  & 0.9 $\mu$b $(10\%)$         & 2  $ (15\%)$    & 1.4 $ (15\%)$  & 1.5    \\
  \hline
$^{252}$Fm  &                        & 0.11 $ (35\%)$ & 0.06 $ (20\%)$ & 0.8    \\
$^{254}$Fm  & 0.9 $\mu$b $(10\%)$         & 0.81 $ (5\%)$  & 0.7 $ (8\%)$   & 1.1    \\
$^{255}$Fm  &                        & 0.9 $ (30\%)$  & 0.62 $ (10\%)$ & 0.47    \\
$^{256}$Fm  & 39 nb $(36\%)$          & 0.24 $ (20\%)$ & 0.14 $ (15\%)$ & 0.07    \\
\hline
  \end{tabular}
\end{spacing}
\end{table}

\begin{figure*}[htb]
\includegraphics[width=1.\linewidth]{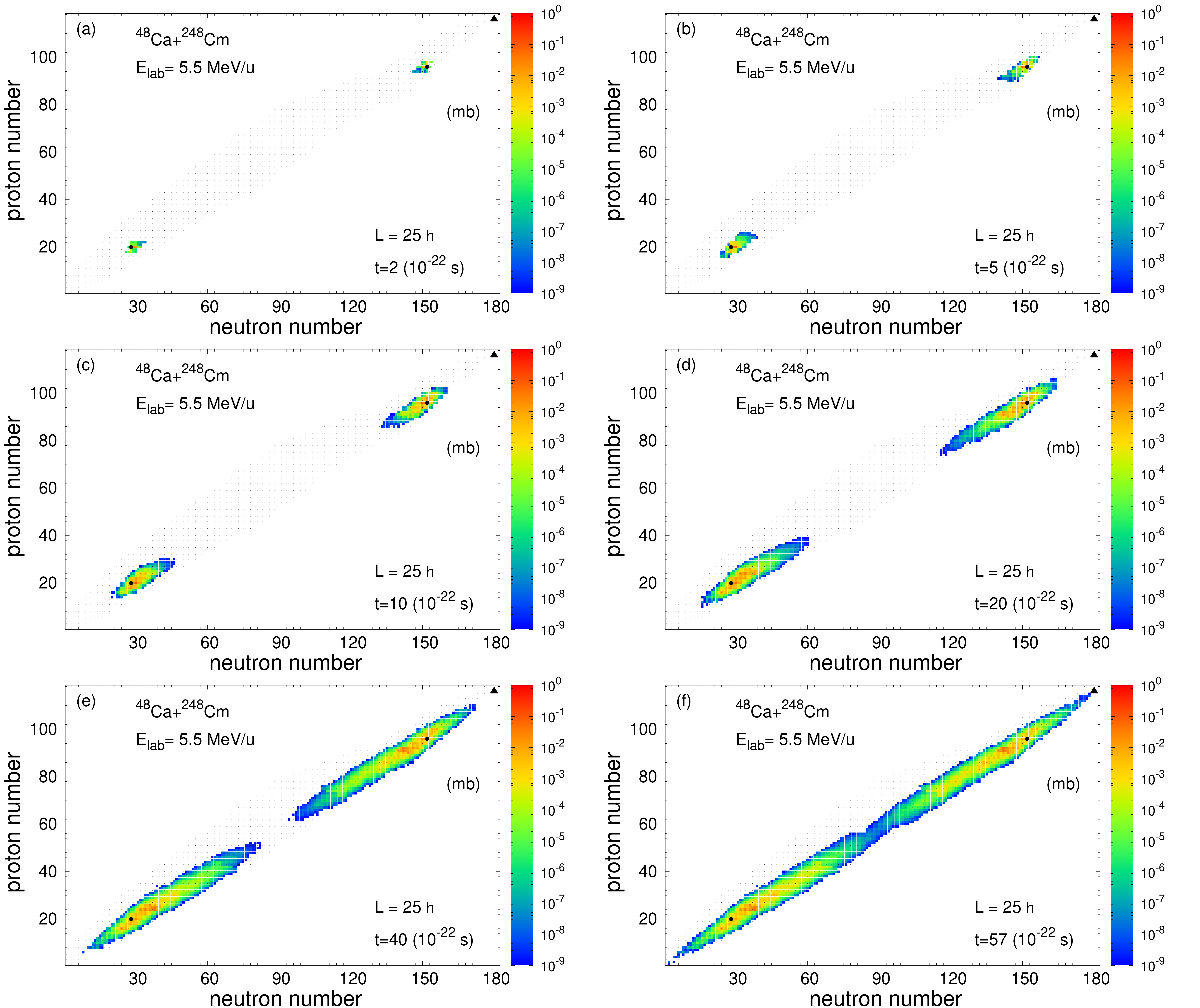}
\caption{\label{fig6}(Color online) The production cross sections of primary fragments in collisions of $^{48}$Ca on $^{248}$Cm at the incident energy of 5.5 MeV/nucleon with the initial angular momentum of  L = 25 $\hbar$. Isotopes yields in panel (a), (b), (c), (d), (e) and (f) correspond to evolution time 2$\times$10$^{-22}$ s, 4$\times$10$^{-22}$ s, 1$\times$10$^{-21}$ s, 2$\times$10$^{-21}$ s, 4$\times$10$^{-21}$ s and 5.7$\times$10$^{-21}$ s, respectively.}
\end{figure*}
Collisions of atomic nuclei are ideal to investigate equilibration and dissipative process in quantum many body system\cite{mor81,vio87}.
Exploring nuclear dynamics in complete and incomplete fusion for heavy ion collisions can be used to understand the interplay between equilibrium and dissipation in quantum system.
In collision process, neucleons can diffuse from target to compound nuclei, where probabilties of all formed fragments will be exported at every moment.
Then, the dynamical process of isotopic yields from pre-equilibrium to equilibrium can reveal a boundary line bewteen complete fusion and multinucleon trnsfer reactions.
The timescale for mass equilibrium ($\sim 10^{-20}$ s) is found to be larger than timescale for kinetic energy dissipations which is on the order of $10^{-21}$ s.
In our approach, collisions of $^{48}$Ca + $^{248}$Cm at $E_{\rm lab}$ = 5.5 MeV/nucleon with impact parameter L = 25 $\hbar$, dynamical nucleon transfer between projectile and target have been exhibited by plotting graphs for all fragments production in different timescales, when they were 2$\times$10$^{-22}$ s, 4$\times$10$^{-22}$ s, 1$\times$10$^{-21}$ s, 2$\times$10$^{-21}$ s, 4$\times$10$^{-21}$ s and 5.7$\times$10$^{-21}$ s, where they located in Fig.\ref{fig6} (a), (b), (c), (d), (e) and (f), respectively. 
One can see that the composite system starts to fusion compound nuclei from 5.7$\times$10$^{-21}$ s. 
We consider the timescale as a boundary between complete and incomplete fusion.
The moments of collision partners fusing to compound nuclei are found in diffusion process. 
We calculate and find all these moments corresponding to all impact parameters. 
Plot all these moments as a red dash line shown in Fig. \ref{fig1}(b). 
It was found that the up limitation of impact parameter for synthsis of superheavy element $Z$ = 116 was $L$ = 56 $\hbar$, mainly because the dissipatting energy of colliding system reached equilibrium almostly. 
The boundary line between CF and MNT have been found,  which is around 5.7$\times 10^{-21}$ s.
It is worth mentioning that our calculations for equilibrium timescales of fragment mass asymmetry and kinetic dissipation are consistant with those calculations from TDHF and TDRPA \cite{sim20}.

\section{Conclusions}\label{sec4}

In summary, production of above-target isotopes and compound nuclei has been investigated within the DNS model through complete fusion-evaporation and multinucleon transfer reactions, for the MNT products of Bk, Cf, Es, Fm, for the fusion-evaporation products of $^{292,293}$Lv. 
In the collision process, kinetic energy dissipating in internal excitation energy heat up the composite system. 
The nucleon transfer takes place at the touching configuration of two colliding nuclei under the PES. The valley shape of the PES influences the formation of primary fragments and leads to the production of quasi-fission isotopes. PES enable here is a diabatic type, which is derived by $Q_{\rm gg}$ value, double-folding nuclear potential and Coulomb potential. 
The TKE-mass distribution of multinucleon trnsfer products reveals some quantities, namely, reaction mechanisms, dissipating energy, shell and structure effect.
The calculation can explain both experiments results of complete fusion-evaporation products and multinucleon transfer fragments for $^{48}$Ca + $^{248}$Cm reasonably. 
The available experiment data are achieved from laboratories all over the world, successively. 
In our calculation, the diffusion pathway from target to compound nuclei has been indicated, derived by dynamical competting with deep-inelastic and quasi fission for two heavy systems. 
The excitation functions of producing superheavy isotopes $^{292,293}$Lv are composed of experiment data from three different laboratories GSI, FLNR and RIKEN. 

We compare their experiment data obtained from two groups, which are taget-like fragments in collision of $^{48}$Ca + $^{248}$Cm, where they have been performed at GSI $\&$ LBL in the year 1986 and 2010, repectively.
It is found that the obtained isotopic cross section are highly dependent on identification method.
In particular, for below-target isotopes ($Z \textless$ 96),  the cross section obtained by radiochemical mothed has three order of magnitude larger than that obtained by decay spectoscopy.
However, the cross section of above-target nuclei ($Z \textgreater$ 96) from both expriments are quite well consistent despite the different experimental techniques and systematic uncertainties.
The effective impact parameter of these two colliding partners leading to compound nuclei is from center collison to $L$ = 52 $\hbar$. The timescale between complete and incomplete reactions is about 5.7$\times 10^{-21}$ s with effective impact parameters. We predict that synthesis cross-sections of unknown rare isotopes $^{254}$Bk, $^{257}$Cf, $^{259}$Es, $^{260}$Fm, $^{263}$Md, $^{264,265}$No, $^{266}$Lr and $^{268}$Rf are around nanobarn in collisions of $^{48}$Ca + $^{248}$Cm nearby Coulomb barrier energies.

\section{Acknowledgements}
This work was supported by National Science Foundation of China (NSFC) (Grants No. 12105241,12175072), NSF of Jiangsu Province (Grants No. BK20210788), Jiangsu Provincial Double-Innovation Doctor Program(Grants No. JSSCBS20211013) and University Science Research Project of Jiangsu Province (Grants No. 21KJB140026). This project was funded by the Key Laboratory of High Precision Nuclear Spectroscopy, Institute of Modern Physics, Chinese Academy of Sciences (CAS). This work was supported by the Strategic Priority Research Program of CAS (Grant No. XDB34010300).


\begin{thebibliography}{99}
\bibitem{og15} Y. T. Oganessian, V. K. Utyonkov, Superheavy nuclei from $^{48}$Ca-induced reactions, \href{https://doi.org/10.1016/j.nuclphysa.2015.07.003}{Nucl. Phys. A  \textbf{944}, 62-98 (2015)}. 
\bibitem{ut15} V. K. Utyonkov, N. T. Brewer, and Yu. Ts. Oganessian et al, Experiments on the synthesis of superheavy nuclei $^{284}$Fl and $^{285}$Fl in the $^{239,240}$Pu+$^{48}$Ca reactions, \href{https://doi.org/10.1103/PhysRevC.92.034609}{Phys. Rev. C \textbf{92}, 034609 (2015)}. 
\bibitem{ho11} S. Hofmann, Radiochim, Synthesis of superheavy elements by cold fusion, 
\href{https://doi.org/10.1524/ract.2011.1854}{Acta. \textbf{99}, 405 (2011)}.
\bibitem{mu15} CE Düllmann, RD Herzberg, W Nazarewicz, et al.,Special Issue on Superheavy Elements-Foreword, \href{https://doi.org/10.1016/j.nuclphysa.2015.11.004}{Nucl. Phys. A \textbf{944}, 5-29 (2015)}.
\bibitem{mo15} K. Morita, SHE research at RIKEN/GARIS, \href{https://doi.org/10.1016/j.nuclphysa.2015.10.007}{ Nucl. Phys. A \textbf{944}, 30-61 (2015)}.
\bibitem{st09} L. Stavsetra, K. E. Gregorich, J. Dvorak, et. al., Independent Verification of Element 114 Production in the $^{48}$Ca + $^{242}$ Pu Reaction, \href{https://doi.org/10.1103/PhysRevLett.103.132502}{Phys. Rev. Lett. \textbf{103}.132502 (2009)}.
\bibitem{ga01} Z. G. Gan, Z. Qin, H. M. Fan, et al., A new alpha-particle-emitting isotope $^{259}$Db, \href{https://doi.org/10.1007/s100500170140}{Euro. Phys. J. A \textbf{10}, 21 (2001)} .
\bibitem{ga04} Z. G. Gan, J. S. Guo, X. L. Wu,et al., New isotope $^{265}$Bh, \href{https://doi.org/10.1140/epja/i2004-10020-2}{Euro. Phys. J. A \textbf{20}, 385 (2004)}.
\bibitem{zh12} Z. Y. Zhang, Z. G. Gan, L. Ma, et. al., Observation of the Superheavy Nuclide $^{271}$Ds, \href{https://doi.org/10.1088/0256-307x/29/1/012502}{Chin. Phys. Lett. \textbf{29(1)}, 012502 (2012)}.
\bibitem{Art71} A. G. Artukh, V. V. Avdeichikov, G. F. Gridnev, V. L. Mikheev, V. V. Volkov and J. Wilczynski, New isotopes $^{29,30}$Mg, $^{31,32,33}$Al, $^{33,34,35,36}$Si, $^{35,36,37,38}$P, $^{39,40}$S and $^{41,42}$Cl produced in bombardment of a $^{232}$Th target with 290 MeV $^{40}$Ar ions, \href{https://doi.org/10.1016/0370-2693(85)91161-X}{Nul. Phys. A \textbf{176}, 284-288 (1971)}.
\bibitem{Art73} A. G. Artukh, G. F. Gridnev, V. L. Mikheev, V. V. Volkov and J. Wilczynski, Multinucleon transfer reactions in the $^{232}$Th + $^{22}$Ne system, \href{https://doi.org/10.1016/0370-2693(85)91161-X}{Nul. Phys. A \textbf{211}, 299-309 (1973)}.
\bibitem{Art74} A. G. Artukh, G. F. Gridnev, V. L. Mikheev, V. V. Volkov and J. Wilczynski, Transfer reactions in the interaction of $^{40}$Ar with $^{232}$Th, \href{https://doi.org/10.1016/0370-2693(85)91161-X}{Nul. Phys. A \textbf{215}, 91-108 (1973)}.
\bibitem{Hil77} K. D. Hildenbrand, H. Freiesleben, F. Phlhofer, W. F. W. Schneider, R. Bock, D. v. Harrach, and H. J. Specht, Reaction between $^{238}$U and$^{ 238}$U at 7.42 MeV Nucleon, \href{https://link.aps.org/doi/10.1103/PhysRevLett.39.1065}{Phys. Rev. Lett. \textbf{39}, 1065 (1977)}.
\bibitem{Gla79}P. Gl\"{a}ssel, D. V. Harrach, Y. Civelekoglu, R. M\"{a}nner, H. J. Specht, J. B. Wilhelmy, H. Freiesleben, and K. D.
Hildenbrand, Three-Particle Exclusive Measurements of the Reactions $^{238}$U+$^{238}$U and $^{238}$U+$^{248}$Cm, \href{https://link.aps.org/doi/10.1103/PhysRevLett.43.1483}{Phys. Rev. Lett. \textbf{ 43}, 1483 (1979)}.
\bibitem{Moo86} K. J. Moody, D. Lee, R. B. Welch, K. E. Gregorich, G. T. Seaborg, R. W. Lougheed, and E. K. Hulet, Actinide production in reactions of heavy ions with $^{248}$Cm, \href{https://link.aps.org/doi/10.1103/PhysRevC.33.1315}{Phys. Rev. C \textbf{33}, 1315 (1986)}.
\bibitem{Wel87} R. B. Welch, K. J. Moody, K. E. Gregorich, D. Lee, and G. T. Seaborg, Dependence of actinide production on the mass number of the projectile: Xe+$^{248}$Cm, \href{https://link.aps.org/doi/10.1103/PhysRevC.35.204}{Phys. Rev. C \textbf{35}, 204 (1987)}.
\bibitem{Ko12} E. M. Kozulin, E. Vardaci, G. N. Knyazheva, A. A. Bogachev, S. N. Dmitriev, I. M. Itkis, M. G. Itkis, A. G. Knyazev, T. A. Loktev, K. V. Novikov, E. A. Razinkov, O. V. Rudakov, S. V. Smirnov, W. Trzaska, and V. I. Zagrebaev, Mass distributions of the system $^{136}$Xe+$^{208}$Pb at laboratory energies around the Coulomb barrier: A candidate reaction for the production of neutron-rich nuclei at N=126, \href{https://doi.org/10.1103/PhysRevC.86.044611}{Phys. Rev. C \textbf{86}, 044611 (2012)}.
\bibitem{Ba15} J. S. Barrett, W. Loveland, R. Yanez \emph{et al.}, $^{136}$Xe+$^{208}$Pb reaction: A test of models of multinucleon transfer reactions,  \href{https://doi.org/10.1103/PhysRevC.91.064615}{Phys. Rev. C \textbf{91}, 064615 (2015)}.
\bibitem{Wa15} Y. X. Watanabe \emph{et al.}, Pathway for the Production of Neutron-Rich Isotopes around the N = 126 Shell Closure, \href{https://doi.org/10.1103/PhysRevLett.115.172503}{Phys. Rev. Lett. \textbf{115}, 172503 (2015)}.
\bibitem{Ko17} E. M. Kozulin, V. I. Zagrebaev, G. N. Knyazheva, I. M. Itkis, K. V. Novikov, M. G. Itkis, S. N. Dmitriev, I. M. Harca, A. E. Bondarchenko, A. V. Karpov, M. G. Itkis, S. N. Dmitriev, I. M. Harca, A. E. Bondarchenko, A. V. Karpov, V. V. Saiko, and E. Vardaci, Inverse quasifission in the reactions $^{156,160}$Gd+$^{186}$W. \href{https://doi.org/10.1103/PhysRevC.96.064621}{Phys. Rev. C \textbf{96}, 064621 (2017)}.
\bibitem{Wu18} S. Wuenschel, K. Hagel, M. Barbui, J. Gauthier, X. G. Cao, R. Wada, E. J. Kim, Z. Majka, R. Planeta, Z. Sosin, A. Wieloch, K. Zelga, S. Kowalski, K. Schmidt, C. Ma, G. Zhang, and J. B. Natowitz, Experimental survey of the production of $\alpha$-decaying heavy elements in $^{238}$U+$^{232}$Th reactions at 7.5-6.1 MeV/nucleon. \href{https://doi.org/10.1103/PhysRevC.97.064602}{Phys. Rev. C \textbf{97}, 064602 (2018)}.
\bibitem{Za07} V. Zagrebaev and W. Greiner. Low-energy collisions of heavy nuclei: dynamics of sticking, mass transfer and fusion, \href{https://doi.org/10.1088/0954-3899/34/1/001}{J. Phys. G \textbf{34}, 1 (2007) }; New way for the production of heavy neutron-rich nuclei, \href{https://doi.org/10.1088/0954-3899/35/12/125103}{J. Phys. G \textbf{35}, 125103 (2008)}.
\bibitem{Za08} V. Zagrebaev and W. Greiner. Synthesis of superheavy nuclei: A search for new production reactions, \href{https://doi.org/10.1103/PhysRevC.78.034610}{ Phys. Rev. C \textbf{78}, 034610 (2008)}; Production of New Heavy Isotopes in Low-Energy Multinucleon Transfer Reactions. \href{https://doi.org/10.1103/PhysRevLett.101.122701}{Phys. Rev. Lett. \textbf{101}, 122701 (2008)}.
\bibitem{Gola09} C. Golabek and C. Simenel, Collision Dynamics of Two $^{238}$U Atomic Nuclei, \href{https://doi.org/10.1103/PhysRevLett.103.042701}{Phys. Rev. Lett. \textbf{103}, 042701 (2009)}.
\bibitem{Seki16} K. Sekizawa and K. Yabana, Time-dependent Hartree-Fock calculations for multinucleon transfer and quasifission processes in the $^{64}$Ni+$^{238}$U reaction, \href{https://doi.org/10.1103/PhysRevC.93.054616}{ Phys. Rev. C \textbf{93}, 054616 (2016)}.
\bibitem{Gu18} L. Guo, C. Simenel, L. Shi, and C. Yu, The role of tensor force in heavy-ion fusion dynamics, \href{https://doi.org/10.1016/j.physletb.2018.05.066}{ Phys. Lett. B \textbf{782}, 401-405 (2018)}.
\bibitem{Ji18} X. Jiang and N. Wang, Production mechanism of neutron-rich nuclei around N = 126 in the multi-nucleon transfer reaction $^{132}$Sn + $^{208}$Pb,  \href{https://doi.org/10.1088/1674-1137/42/10/104105}{Chin. Phys. C \textbf{42}, 104105 (2018)}.
\bibitem{Wint94}  A. Winther, Grazing reactions in collisions between heavy nuclei, \href{https://doi.org/10.1016/0375-9474(94)90430-8} {Nucl. Phys. A \textbf{572},191 (1994)}; Dissipation, polarization and fluctuation in grazing heavy-ion collisions and the boundary to the chaotic regime, \href{https://doi.org/10.1016/0375-9474(95)00374-A}{ Nucl. Phys. A \textbf{594}, 203 (1995)}.
\bibitem{bib:6} \href{http://www.to.infn.it/nanni/grazing}{http://www.to.infn.it/nanni/grazing.}
\bibitem{Zha15} K. Zhao, Z. Li, N. Wang, Y. Zhang, Q. Li, Y. Wang, and X. Wu, Production mechnasim of neutron-rich transuranium nuclei in $^{238}$U+$^{238}$U. \href{https://doi.org/10.1103/PhysRevC.92.024613}{ Phys. Rev. C \textbf{92}, 024613 (2015)}.
\bibitem{li16} C. Li, F. Zhang, J. J. Li, L. Zhu, J. L. Tian, N. Wang, and F. S. Zhang, Multinucleon transfer in the $^{136}$Xe+$^{208}$Pb reaction, \href{https://doi.org/10.1103/PhysRevC.93.014618}{Phys. Rev. C \textbf{93}, 014618 (2016)}.
\bibitem{sh11} C. W. Shen, D. Boilley, Q. F. Li, J. J. Shen, and Y. Abe, Fusion hindrance in reactions with very heavy ions: Border between normal and hindered fusion, \href{https://doi.org/10.1103/PhysRevC.83.054620}{Phys. Rev. C \textbf{83}, 054620 (2011)}.
\bibitem{bo11} D. Boilley, H. L. Lü, C. W. Shen, Y. Abe, and B. G. Giraud, Fusion hindrance of heavy ions: Role of the neck, \href{https://doi.org/10.1103/PhysRevC.84.054608}{Phys. Rev. C, \textbf{84}, 054608 (2011)}.
\bibitem{Fe09} Z. Q. Feng, G. M. Jin, and J. Q. Li, Production of heavy isotopes in transfer reactions by collisions of $^{238}$U+$^{238}$U, \href{https://doi.org/10.1103/PhysRevC.80.067601}{Phys. Rev. C \textbf{80}, 067601 (2009)}.
\bibitem{bib:8} G. G. Adamian, N. V. Antonenko, V. V. Sargsyan, and W. Scheid, Possibility of production of neutron-rich Zn and Ge isotopes in multinucleon transfer reactions at low energies. \href{https://doi.org/10.1103/PhysRevC.81.024604}{Phys. Rev. C \textbf{81}, 024604 (2010)}; Predicted yields of new neutron-rich isotopes of nuclei with Z=64-80 in the multinucleon transfer reaction $^{48}$Ca+$^{238}$U, \href{https://doi.org/10.1103/PhysRevC.81.057602}{Phys. Rev. C \textbf{81}, 057602 (2010)}.
\bibitem{ch20} P. H. Chen, F. Niu, W. Zuo, and Z. Q. Feng, Approaching the neutron-rich heavy and superheavy nuclei by multinucleon transfer reactions with radioactive isotopes,  \href{https://journals.aps.org/prc/abstract/10.1103/PhysRevC.101.024610}{Phys. Rev. C, \textbf{101}: 024610, (2020)}.
\bibitem{bao18} X. J. Bao, S. Q. Guo, H. F. Zhang, and J. Q. Li, Dynamics of complete and incomplete fusion in heavy ion collisions,  \href{https://doi.org/10.1103/PhysRevC.97.024617}{Phys. Rev. C, \textbf{100}: 024617, (2018)}.
\bibitem{ba21} X. J. Bao, Possibilities for synthesis of new transfermium isotopes in multinucleon transfer reactions, \href{https://doi.org/10.1103/PhysRevC.104.034604}{Phys. Rev. C \textbf{104}, 034604 (2021)}.
\bibitem{zhu21} L. Zhu, Shell inhibition on production of N = 126 isotones in multinucleon transfer reactions, \href{https://doi.org/10.1016/j.physletb.2021.136226} {Phys. Lett. B \textbf{816}, 136226 (2021)}.
\bibitem{og00} Yu. Ts. Oganessian, V. K. Utyonkov, Yu. V. Lobanov, Observation of the decay of $^{292}$116, 
\href{https://doi.org/10.1103/PhysRevC.63.011301}{Phys. Rev. C, \textbf{63}, 011301(R), (2000)} .
\bibitem{ho12} S. Hofmann, S. Heinz, R. Mann et. al., The reaction $^{48}$Ca + $^{248}$Cm $\rightarrow$ $^{296}$116$^*$ studied at the GSI-SHIP, \href{https://doi.org/10.1140/epja/i2012-12062-1}{Eur. Phys. J. A \textbf{48}, 62 (2012)}.
\bibitem{ka17} D. Kaji, K. Morita1, K. Morimoto et. al., Study of the Reaction $^{48}$Ca + $^{248}$Cm $\rightarrow$ $^{296}$116$^*$ at RIKEN-GARIS, \href{https://doi.org/10.7566/JPSJ.86.034201}{Journal of the Physical Society of Japan \textbf{86}, 034201 (2017)}.
\bibitem{Deva15} H. M. Devaraja, S. Heinz, O. Beliuskina, V. Comas, S. Hofmann, C. Hornung, G. Mnzenberg, K. Nishio, D. Ackermann, Y.K. Gambhir, M. Gupta, R.A. Henderson, F.P. Heberger, J. Khuyagbaatar, B. Kindler, B. Lommel, K.J. Moody, J. Maurer, R. Mann, A.G. Popeko, D.A. Shaughnessy, M.A. Stoyer, A.V. Yeremin, Observation of new neutron-deficient isotopes with Z $\ge$ 92 in multinucleon transfer reactions, \href{http://dx.doi.org/10.1016/j.physletb.2015.07.006}{ Phys. Lett. B \textbf{748}, 199-203 (2015)}.
\bibitem{hof85} D. C. Hoffman, M. M. Fowler, and W. R. Daniels et. al., Excitation functions for production of heavy actinides from interactions of $^{40}$Ca and $^{48}$Ca ions with $^{248}$Cm, \href{https://doi.org/10.1103/PhysRevC.31.1763}{Phys. Rev. C, \textbf{31}, 1763 (1985)}.
\bibitem{gag86} H. Giggeler, W. Bruchle, M. Brugger et. al., Production of cold target-like fragments in the reaction of $^{48}$Ca + $^{248}$Cm, \href{https://doi.org/10.1103/PhysRevC.33.1983}{Phys. Rev. C, \textbf{33}, 1983 (1986)}.
\bibitem{hez16} S. Heinz, H.M. Devaraja, O. Beliuskina et. al., Synthesis of new transuranium isotopes in multinucleon transfer reactions using a velocity filter, \href{https://doi.org/10.1140/epja/i2016-16278-7}{Eur. Phys. J. A  \textbf{52}, 278 (2016)}.
\bibitem{dev19} H. M. Devaraja, S. Heinz, O. Beliuskina et. al., Population of nuclides with Z $\leq$ 98 in multi-nucleon transfer reactions of $^{48}$Ca + $^{248}$Cm, \href{https://doi.org/10.1140/epja/i2019-12737-y}{Eur. Phys. J. A \textbf{55}, 25(2019)}.
\bibitem{Ch17} P. H. Chen, Z. Q. Feng, J. Q. Li, and H. F. Zhang, Production of proton-rich nuclei around Z = 84-90 in fusion-evaporation reactions, \href{https://doi.org/10.1140/epja/i2017-12281-x}{Eur. Phys. J. A \textbf{53}, 95 (2017)}.
\bibitem{No75} W. N\"{o}renberg, Quantum-statistical approach to gross properties of peripheral collisions between heavy nuclei, \href{https://link.springer.com/article/10.1007/BF01437736}{Z. Phys. A \textbf{274}, 241 (1975)}.
\bibitem{Wo78} G. Wolschin and W. N\"{o}renberg, Analysis of relaxation phenomena in heavy-ion collisions \href{https://link.springer.com/article/10.1007/BF01411331}{Z. Phys. A \textbf{284}, 209 (1978)}.
\bibitem{Fe07b} Z. Q. Feng, G. M. Jin, F. Fu, J. Q. Li, Isotopic dependence of production cross sections of superheavy nuclei in hot fusion reactions, \href{http://hepnp.ihep.ac.cn/article/id/02c1a9d3-a94a-4eb1-8f6d-6659479c5784}{Chin. Phys. C 31, 366 (2007)}.
\bibitem{Ch16} P. H. Chen, Z. Q. Feng, J. Q. Li, and H. F. Zhang, A statistical approach to describe highly excited heavy and superheavy nuclei, \href{https://iopscience.iop.org/article/10.1088/1674-1137/40/9/091002/meta}{Chin. Phys. C \textbf{40}, 091002 (2016)}.
\bibitem{Fe06} Z. Q. Feng, G. M. Jin, F. Fu, and J. Q. Li, Production cross sections of superheavy nuclei based on dinuclear system model, \href{https://doi.org/10.1016/j.nuclphysa.2006.03.002}{Nucl. Phys. A \textbf{771}, 50 (2006)}; Z. Q. Feng, G. M. Jin, J. Q. Li, and W. Scheid, Formation of superheavy nuclei in cold fusion reactions, \href{https://doi.org/10.1103/PhysRevC.76.044606}{Phys. Rev. C \textbf{76}, 044606 (2007)}.
\bibitem{itk04} M. G. Itkis, J. Aysto, S. Beghini, et al., Shell effects in fission and quasi-fission of heavy and superheavy nuclei, \href{https://doi.org/10.1016/j.nuclphysa.2004.01.022} {Nucl. Phys. A \textbf{734}, 136-147 (2004)}.
\bibitem{zag05} V. Zagrebaev and W. Greiner, Unified consideration of deep inelastic, quasi-fission and fusion-fission phenomena,\href{doi:10.1088/0954-3899/31/7/024}
{J. Phys. G: Nucl. Part. Phys. \textbf{31},  825–844 (2005)}.
\bibitem{oga04} Yu. Ts. Oganessian, et al., Measurements of cross sections and decay properties of the isotopes of elements 112, 114, and 116 produced in the fusion reactions $^{233,238}$U, $^{242}$Pu, and  
$^{248}$Cm + $^{48}$Ca, \href{https://doi.org/10.1103/PhysRevC.70.064609} {Phys. Rev. C \textbf{70}, 064609 (2004)}.
\bibitem{zhi21} Z. Cheng and X. J. Bao, Formation of heavy neutron-rich nuclei by $^{48}$Ca-induced multinucleon transfer reactions, \href{https://doi.org/10.1103/PhysRevC.103.024613}{Phys. Rev. C \textbf{103}, 024613 (2021)}.
\bibitem{mor81} L. G. Moretto and R. P. Schmitt, Deep inelastic reactions: A probe of the collective properties of nuclear matter, \href{https://doi.org/10.1088/0034-4885/44/5/002}{Rep. Prog. Phys. \textbf{44}, 533 (1981)}.
\bibitem{vio87} V. E. Viola, Nucleus-nucleus collisions: A laboratory for studying equilibration phenomena, \href{https://doi.org/10.1021/ar00133a005}{Acc. Chem. Res. \textbf{20}, 32 (1987)}.
\bibitem{sim20} C. Simenel, K. Godbey, and A. S. Umar, Timescales of Quantum Equilibration, Dissipation and Fluctuation in Nuclear Collisions, \href{https://doi.org/10.1103/PhysRevLett.124.212504} {Phys. Rev. Lett. \textbf{124}, 212504 (2020)}.
\end{thebibliography}
\end{document}